\documentclass[superscriptaddress,amsmath, nofootinbib]{revtex4}
\usepackage{amsfonts}
\usepackage{graphicx}
\usepackage[brazil]{babel}
\usepackage[latin1]{inputenc}
\usepackage{amsmath}
\usepackage{amssymb}
\begin{document}


\title{General solution to the Euler-Poisson equations of a free Lagrange top directly for the rotation matrix.}

\author{Alexei A. Deriglazov }
\email{alexei.deriglazov@ufjf.br} \affiliation{Depto. de Matem\'atica, ICE, Universidade Federal de Juiz de Fora,
MG, Brazil} 


\date{\today}

\begin{abstract}
The Euler-Poisson equations para determinar the rotation matrix of a rigid body can be solved without using of particular parameterization like the Euler angles. For the free Lagrange top, we obtain and discuss a general analytic solution, and compare it with the Poinsot's picture of motion. 
\end{abstract}

\maketitle 



\section{Introduction.}

This paper is the fourth in the series \cite{AAD23, AAD23_1,AAD23_2}, devoted to a systematic exposition of the dynamics of a rigid body, considered as  a system with kinematic constraints. The constraints can be taken into account using an appropriate Lagrangian action (see Eq. (12) in \cite{AAD23}). 
Having accepted this expression, we no longer need any additional postulates or assumptions about the behavior of the rigid body. As was shown in \cite{AAD23}, all the basic quantities and characteristics of a rigid body, as well as the equations of motion and integrals of motion, are obtained from the variational problem by direct and unequivocal calculations within the framework of standard methods of classical mechanics. In particular, for the rotational degrees of freedom this variational problem lead to the Euler-Poisson equations
\begin{eqnarray}
I\dot{\boldsymbol\Omega}=[I{\boldsymbol\Omega}, {\boldsymbol\Omega}], \label{s0} \\   
\dot R_{ij}=-\epsilon_{jkm}\Omega_k R_{im},  \label{s1}
\end{eqnarray}
that should be solved with universal initial conditions for the rotation matrix: $R_{ij}(0)=\delta_{ij}$. The initial conditions for $\Omega_i$ can be any three numbers, they represent the initial velocity of rotation of the body. These equations represent the Hamiltonian system for mutually independent variables $R_{ij}(t)$ and $\Omega_i(t)$ \cite{AAD23}. Here $R_{ij}(t)$ is $3\times 3$\,-matrix while $\Omega_i(t)$ is the Hamiltonian counterpart of angular velocity in the body. The inertia tensor $I$ is assumed to be diagonal: $I=diagonal (I_1, I_2 I_3)$.  Given the solution $R(t)$, the evolution of the body's point ${\bf y}(t)$ is restored according to the rule: ${\bf y}(t)={\bf C}_0+{\bf V}_0 t+R(t){\bf x}(0)$, where ${\bf x}(0)$ is the initial position of the point in the center-of-mass system, while  the term ${\bf C}_0+{\bf V}_0 t$ describes the motion of the center of mass with respect to the Laboratory. Recall also that both columns and rows of the matrix $R_{ij}$ have a simple interpretation. The columns $R(t)=({\bf R}_1, {\bf R}_2, {\bf R}_3)$ form an orthonormal basis rigidly connected to the body. The rows $R^T(t)=({\bf G}_1, {\bf G}_2, {\bf G}_3)$ represent the laboratory basis vectors ${\bf e}_i$ in the rigid body basis. For example, the numbers ${\bf G}_1(t)=(R_{11}, R_{12}, R_{13})$ are components of the basis vector ${\bf e}_1$ in the basis ${\bf R}_i(t)$. 

The Euler-Poisson equations admite various  integrals of motion. They are the rotational energy, 
\begin{eqnarray}\label{s1.1}
2E=(\boldsymbol{\Omega} I \boldsymbol{\Omega})=I_1\Omega_1^2+I_2\Omega_2^2+I_3\Omega_3^2, 
\end{eqnarray}
three components of angular momentum 
\begin{eqnarray}\label{s1.2}
m_i=(RI\boldsymbol{\Omega})_i=I_1R_{i1}\Omega_1+I_2R_{i2}\Omega_2+I_3R_{i3}\Omega_3,  
\end{eqnarray}
and six more integrals 
\begin{eqnarray}\label{s1.2.0}
R_{ki}R_{kj}=a_{ij}. 
\end{eqnarray}
The integration constants $a_{ij}$ are fixed by the initial conditions: $R_{ij}(0)=\delta_{ij}$ implies $a_{ij}=\delta_{ij}$. Due to this, any  solution $R_{ij}(t)$ to the system with these initial conditions authomatically will be the orthogonal matrix at any future instant: $R^T(t)R(t)=1$.  Besides, the initial conditions imply the following relation between the energy and angular momentum: 
\begin{eqnarray}\label{s1.3}
2E=\frac{1}{I_1}m_1^2+\frac{1}{I_2}m_2^2+\frac{1}{I_3}m_3^2. 
\end{eqnarray}
The square of angular momentum does not contain $R_{ij}$
\begin{eqnarray}\label{s1.4}
{\bf m}^2=I_1^2\Omega_1^2+I_2^2\Omega_2^2+I_3^2\Omega_3^2,
\end{eqnarray}
so the Euler equations itself admite two independent integrals of motion (\ref{s1.1}) and  (\ref{s1.4}). 

The equations (\ref{s1})  are  written for an excess number of variables. Indeed, at each instant of time, the nine matrix elements $R_{ij}$ obey to six constraints $R^TR =1$, so we need to know only some $9-6=3$ independent parameters to specify the matrix $R$. 
It becomes a centenary tradition in the text-books to discuss solutions to these equations using some irreducible set of variables like the Euler angles \cite{Whit_1917,Mac_1936,Lei_1965}. However, there are a number of arguments against this way of presentation. 
First, in the discussions based on the Euler angles, the equations (\ref{s1}) are so dissolved in the calculations that sometimes even not mentioned. 
Second, to describe a rigid body, we need to know namely the evolution of $R_{ij}(t)$. If so, why do we then insist on introducing independent variables? 
Third, the description in terms of independent variables often turns out to be local, which can lead to misunderstandings, see \cite{AAD23_2}. Finally, solving the Euler-Poisson equations  directly for the original variables sometimes requires less effort than solving the same equations through independent variables, see for example the solution to Euler equations for the free asymmetric top in \cite{Landau_8}. 

Here we solve the Euler-Poisson equations for the free Lagrange top directly for the variables $R_{ij}(t)$, without using any parametrization, and present the general solution for $R_{ij}(t)$ through the trigonometric functions. 

Using the lines ${\bf G}_1, {\bf G}_2$ and ${\bf G}_3$ of the matrix $R$, the system (\ref{s1}) reads as follows 
$I\dot{\boldsymbol\Omega}=[I{\boldsymbol\Omega}, {\boldsymbol\Omega}]$,  $\dot{\bf G}_i=[{\bf G}_i, {\boldsymbol\Omega}]$. Therefore, the original system splits into three. It is sufficient to solve only one system of six equations, that consist of three Euler equations and three equations for any one among the vectors ${\bf G}_i$, 
say, ${\boldsymbol\gamma}$:  
\begin{eqnarray}\label{s2}
I\dot{\boldsymbol\Omega}=[I{\boldsymbol\Omega}, {\boldsymbol\Omega}], 
\end{eqnarray}
\begin{eqnarray}\label{s2.1} 
\dot{\boldsymbol\gamma}=[{\boldsymbol\gamma}, {\boldsymbol\Omega}].  
\end{eqnarray}
Then the whole rotation matrix can be restored, choosing three particular solutions to this system: ${\bf G}_1(t)$ is the solution ${\boldsymbol\gamma}(t)$ that obeys the initial condition ${\boldsymbol\gamma}(0)=(1, 0, 0)$, ${\bf G}_2(t)$ is the solution ${\boldsymbol\gamma}(t)$ that obeys the initial condition ${\boldsymbol\gamma}(0)=(0, 1, 0)$, and ${\bf G}_3(t)$ is the solution ${\boldsymbol\gamma}(t)$ that obeys the initial condition ${\boldsymbol\gamma}(0)=(0, 0, 1)$. 

As we saw above, the Euler equations have two integrals of motion (\ref{s1.1}) and  (\ref{s1.4}). 
The equations (\ref{s2.1}) also have two integrals of  motion following from (\ref{s1.2}) and  (\ref{s1.2.0}): 
\begin{eqnarray}\label{s2.3}
{\boldsymbol\gamma}^2=1, \qquad  c=I_1\Omega_1\gamma_1+I_2\Omega_2\gamma_2+I_3\Omega_3\gamma_3, 
\end{eqnarray}
where $c=m_1$ for ${\boldsymbol\gamma}={\bf G}_1$, and so on.  Using them, the system (\ref{s2}), (\ref{s2.1}) reduces to two first-order equations, each for its own variable. In the case of asymmetric body, they can be formally integrated and give the answer in terms of elliptic 
integrals \cite{Landau_8}.  But in the case of Lagrange (symmetric) top there is an analytic solution for $R_{ij}$ that we present and discuss below. 

For the latter use, we present the relationship among the basic quantities used in the description of a rigid body  (angular velocity $\omega_i$, angular velocity in the body $\Omega_i$, angular momentum $m_i$, and angular momentum in the body $M_i$)  
\begin{eqnarray}\label{s2.4}
{\bf\Omega}=R^T{\boldsymbol\omega}=I^{-1}R^T{\bf m}=I^{-1}{\bf M}. 
\end{eqnarray}

\section{Symmetric top.}

The Lagrange (symmetric) top is a rigid body with two coinciding moments of inertia. So we consider the equations (\ref{s2})-(\ref{s2.3}) with $I_1=I_2$ and with the conserved angular momentum ${\bf m}=(m_1, m_2, m_3)$. First, we confirm that without loss of generality we can assume that $m_1=0$. 

The moments of inertia are eigenvalues of the inertia tensor $I$, with eigenvectors being the body fixed axes at $t=0$: $I{\bf R}_i(0)=I_i{\bf R}_i(0)$.  With $I_1=I_2$ we have $I{\bf R}_1(0)=I_1{\bf R}_1(0)$ and $I{\bf R}_2(0)=I_1{\bf R}_2(0)$, then any linear combination $\alpha{\bf R}_1(0)+\beta{\bf R}_2(0)$ also represents an eigenvector with eigenvalue $I_1$. This means that we are free to choose any two orthogonal axes on the plane ${\bf R}_1(0), {\bf R}_2(0)$ as the inertia axes. We recall that at $t=0$, the Laboratory axes should be chosen along ${\bf R}_i$, since only in this case equations of motion contain $I_i$ instead of the tensor $I_{ij}$. Hence, in the case $I_1=I_2$ we can rotate the Laboratory axes in the plane $(x^1, x^2)$  without breaking the diagonal form of the inertia tensor. Using this freedom, we can assume that $m_1=0$ for our problem. 

{\bf General solution to the Euler equations of a symmetric top.}  With $I_1=I_2$, the Euler equations read
\begin{eqnarray}\label{s3}
\dot\Omega_1=\frac{(I_1-I_3)\Omega_3}{I_1}\Omega_2, \qquad \dot\Omega_2=-\frac{(I_1-I_3)\Omega_3}{I_1}\Omega_1, \qquad \dot\Omega_3=0,
\end{eqnarray}
or
\begin{eqnarray}\label{s4}
\dot\Omega_1=\omega\Omega_2, \qquad \dot\Omega_2=-\omega\Omega_1, \qquad \Omega_3=const,
\end{eqnarray}
where $\phi\equiv(I_1-I_3)\Omega_3/I_1$.  
Their general solution is
$\Omega_1=a\sin(\phi t+\phi_0)$,  $\Omega_2=a\cos(\phi t+\phi_0)$, $\Omega_3=\mbox{const}$.
The solution has a simple meaning: for an observer of the rigid-body frame ${\bf R}_i$, the angular velocity vector ${\boldsymbol\omega}$ precess around ${\bf R}_3$\,-axis with angular frequency $\phi$. 

Let us relate the integration constants $a$, $\Omega_3$ and $\phi_0$ with the basic quantities of a rigid body. The conserved angular 
momentum $m_i$ and the angular velocity $\Omega_i$ are related as follows: $m_i=I_{ij}\Omega_j(0)$, so $m_1=I_1a\sin\phi_0$, $m_2=I_1a\cos\phi_0$,
$m_3=I_3\Omega_3$. Our choice $m_1=0$  fixes the constant $\phi_0=0$ in the equations (\ref{s5}). Then $a=m_2/I_1$ and $\Omega_3=m_3/I_3$. In the result, the first oscillation frequency in the problem is determined by third component of conserved angular momentum: $\phi=(I_1-I_3)m_3/I_1 I_3$.   Taking all this into account, the solution is 
\begin{eqnarray}\label{s5}
\Omega_1=\frac{m_2}{I_1}\sin \phi t, \qquad \Omega_2=\frac{m_2}{I_1}\cos\phi t, \qquad \Omega_3=\frac{m_3}{I_3}.
\end{eqnarray}
The explicit form of the solution shows the conservation of the rotational energy: $2E=I_i\Omega_i^2(t)=m_2^2/I_1+m_3^2/I_3$.

{\bf General solution to the Poisson equations of a symmetric top.} Take the linear integral of motion from Eq. (\ref{s2.3}) and the third component of the Poisson equation
\begin{eqnarray}\label{s6}
\Omega_2\gamma_1-\Omega_1\gamma_2=\dot\gamma_3, \qquad I_1\Omega_1\gamma_1+I_2\Omega_2\gamma_2=c-I_3\Omega_3\gamma_3.
\end{eqnarray}
The solution to this linear system is
\begin{eqnarray}\label{s7}
\gamma_1=\frac{I_1}{m_2^2}\left[I_1\Omega_2\dot\gamma_3+(c-m_3\gamma_3)\Omega_1\right], \qquad 
\gamma_2=\frac{I_1}{m_2^2}\left[-I_1\Omega_1\dot\gamma_3+(c-m_3\gamma_3)\Omega_2\right].
\end{eqnarray}
Substituting these expressions into the Poisson equation $\dot\gamma_1=\Omega_3\gamma_2-\Omega_2\gamma_3$ we get closed equation for $\gamma_3$
\begin{eqnarray}\label{s8}
\ddot\gamma_3+k^2\gamma_3=\frac{m_3}{I_2} c, \qquad 
\mbox{where} \quad k^2\equiv\frac{m_2^2+m_3^2}{I_1}=\frac{{\bf m}^2}{I_1^2}, 
\end{eqnarray}
with the general solution 
\begin{eqnarray}\label{s9}
\gamma_3=b\cos(kt+k_0)+\frac{m_3 c}{{\bf m}^2},  \qquad 
\mbox{where} \quad  k\equiv\frac{|{\bf m}|}{I_1}
\end{eqnarray}
According to Eq. (\ref{s9}), the second oscillation frequency $k$ in the problem is determined by magnitude of conserved angular momentum. Using
this $\gamma_3$ in equations (\ref{s7}), we get the general solution to the system (\ref{s2}), (\ref{s2.1}) as follows
\begin{eqnarray}\label{s10}
\gamma_1=-\frac{|{\bf m}|b}{m_2}\cos\phi t\sin(kt+k_0)+\left[\frac{m_2 c}{{\bf m}^2}-\frac{m_3 b}{m_2}\cos(kt+k_0)\right]\sin\phi t, \cr
\gamma_2=\frac{|{\bf m}|b}{m_2}\sin\phi t\sin(kt+k_0)+\left[\frac{m_2 c}{{\bf m}^2}-\frac{m_3 b}{m_2}\cos(kt+k_0)\right]\cos\phi t, \cr 
\gamma_3=b\cos(kt+k_0)+\frac{m_3 c}{{\bf m}^2}. \qquad \qquad \qquad \qquad \qquad \qquad \qquad \quad 
\end{eqnarray}
At the initial instant $t=0$ we get
\begin{eqnarray}\label{s11}
\gamma_1(0)=-\frac{|{\bf m}|b}{m_2}\sin k_0,  \quad
\gamma_2(0)=\frac{m_2 c}{{\bf m}^2}-\frac{m_3 b}{m_2}\cos k_0, \quad
\gamma_3(0)=b\cos k_0 +\frac{m_3 c}{{\bf m}^2}. 
\end{eqnarray}
As we saw above, the three rows of the matrix $R_{ij}$ are obtained from Eqs. (\ref{s10}) if we choose the integration constants $b$ and $k_0$ so 
that ${\boldsymbol\gamma}(0)=(1, 0, 0)$ with $c=m_1$, then ${\boldsymbol\gamma}(0)=(0, 1, 0)$ with $c=m_2$, and at last ${\boldsymbol\gamma}(0)=(0, 0, 1)$ with $c=m_3$. Solving the equations (\ref{s11}) with these dates, we get, in each case 
\begin{eqnarray}\label{s12}
c=m_1=0, \quad b=-\frac{m_2}{|{\bf m}|}, \quad k_0=\frac{\pi}{2}; \quad
c=m_2, \quad b=-\frac{m_2 m_3}{{\bf m}^2}, \quad k_0=0; \quad
c=m_3, \quad b=\frac{m_2^2}{{\bf m}^2}, \quad k_0=0. 
\end{eqnarray}
Substituting these values into Eq. (\ref{s11})  we get final form of the rotation matrix $R$ of the free Lagrange top as follows:
\begin{eqnarray}\label{s14}
\left(
\begin{array}{ccc}
\cos kt\cos\phi t-\hat m_3\sin kt\sin\phi t & -\cos kt\sin\phi t-\hat m_3\sin kt\cos\phi t  &  \hat m_2\sin kt  \\
{} & {} & {} \\
\hat m_3 \sin kt\cos\phi t+(\hat m_2^2 +\hat m_3^2\cos kt)\sin\phi t & 
-\hat m_3 \sin kt\sin\phi t+(\hat m_2^2 +\hat m_3^2\cos kt)\cos\phi t & \hat m_2\hat m_3(1-\cos kt) \\
{} & {} & {} \\
-\hat m_2 \sin kt\cos\phi t+\hat m_2 \hat m_3(1-\cos kt)\sin\phi t &
\hat m_2 \sin kt\sin\phi t+\hat m_2 \hat m_3(1-\cos kt)\cos\phi t & \hat m_3^2 +\hat m_2^2\cos kt
\end{array}\right) 
\end{eqnarray}
where, assuming $|{\bf m}|\ne 0$, we denoted  by $\hat m_i=m_i/|{\bf m}|$ the components of unit vector in the direction of conserved angular momentum. The two frequences in the problem are $\phi=\frac{I_1-I_3}{I_1 I_3}m_3$ and $k=|{\bf m}|/I_1=\sqrt{m_2^2+m_3^2}/I_1$. Curiously enough, the matrix $R$ depends on the moments of inertia only through the frequences.

\section {Decomposition of the rotation matrix on two subsequent rotations and Poinsot's picture of motion.} 

We start the discussion of the rotation matrix from various limiting cases.

{\bf (1)} Consider the conserved angular momentum ${\bf m}$ in the direction of laboratory axis $OY$, that is $m_3=0$, $m_2\ne 0$. This 
implies $\phi=0$, $|{\bf m}|=|m_2|$, $\hat m_2=\pm 1$, and $\hat m_2\sin |{\bf m}|t/I_1=\sin m_2 t/I_1$. Using this in Eq. (\ref{s14}) we get
\begin{eqnarray}\label{s15}
R_{OY}(t)=\left(
\begin{array}{ccc}
\cos \frac{m_2t}{I_1} & 0 & \sin \frac{m_2t}{I_1} \\ 
0 & 1 & 0 \\
-\sin \frac{m_2t}{I_1} & 0 & \cos \frac{m_2t}{I_1} 
\end{array}\right).  
\end{eqnarray}
This single-frequency motion represents the rotation of the rigid body around the inertia axis $OY$. 

{\bf (2)} Similarly, taking $m_2=0$, $m_3\ne 0$ we get $|{\bf m}|=|m_3|$, $\hat m_3=\pm 1$, $k=|m_3|/I_1$ and $\phi+m_3/I_1=m_3/I_3$. 
Using this in 
Eq. (\ref{s14}) we observe that two frequences are combined into one as follows
\begin{eqnarray}\label{s16}
R_{OZ}(t)=\left(
\begin{array}{ccc}
\cos \frac{m_3t}{I_3}  & -\sin \frac{m_3t}{I_3} & 0  \\ 
\sin \frac{m_3t}{I_3} &  \cos \frac{m_3t}{I_3} & 0 \\
0 & 0 & 1 \\
\end{array}\right).  
\end{eqnarray}
This single-frequency motion represents the rotation of the rigid body around the inertia axis $OZ$. 

{\bf (3)} For the totally symmetric body $I_1=I_2=I_3$ we get $\phi=0$, and the rotation matrix (\ref{s14}) acquires the following form: 
\begin{eqnarray}\label{s17}
R_{\bf m}(t)=\left(
\begin{array}{ccc}
\cos kt & -\hat m_3\sin kt  &  \hat m_2\sin kt  \\
{} & {} & {} \\
\hat m_3 \sin kt & 
\hat m_2^2 +\hat m_3^2\cos kt & \hat m_2\hat m_3(1-\cos kt) \\
{} & {} & {} \\
-\hat m_2 \sin kt & \hat m_2 \hat m_3(1-\cos kt) & \hat m_3^2 +\hat m_2^2\cos kt
\end{array}\right). 
\end{eqnarray}
where $k=|{\bf m}|/I_1$.  The points of the body lying along the axis ${\bf m}$ remain at rest during this movement: for any $c\in {\mathbb R}$ we get $R_{\bf m}(t)c{\bf m}=c{\bf m}$. So this motion represents the rotation of the body around the conserved angular momentum vector ${\bf m}$. This is the only possible motion of totally symmetric body. 

\begin{figure}[t] \centering
\includegraphics[width=12cm]{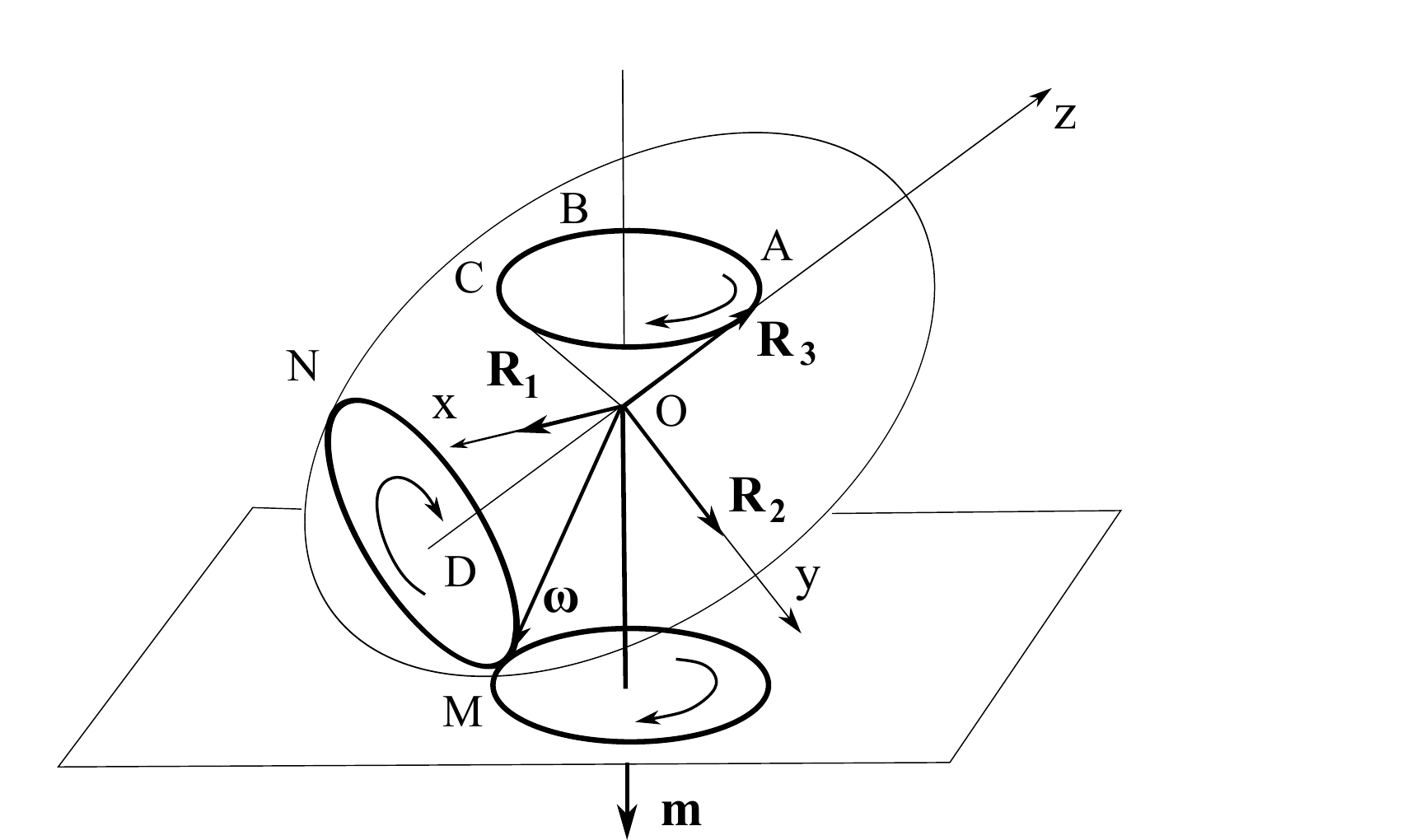}
\caption{Poinsot's picture for the Lagrange top.}\label{TvT_8}
\end{figure}

{\bf (4)} To discuss the general case of two-frequency motion (\ref{s14}), we recall how the rotational motions of the rigid body look like in the picture of Poinsot \cite{Poin,Mac_1936,Lei_1965,AAD23}.  Let's put 
\begin{eqnarray}\label{s18}
I_1=I_2> I_3, \qquad  m_1=0, \qquad m_2>0, \qquad m_3<0. 
\end{eqnarray}
This determines the energy $E=\frac12\sum_i I^{-1}_im_i^2$, the semiaxes $a_i=\sqrt{2E/I_i}$, and the initial angular 
velocity $\omega_i(0)=I^{-1}_im_i$. With these dates, we can construct the invariable plane and the Poinsot's ellipsoid. 
Position of the Poinsot's elipsoid at the initial moment $t=0$ is shown in the Figure \ref{TvT_8}. 
At this moment the vectors ${\bf m}$, ${\boldsymbol\omega}$,  ${\bf R}_2$, ${\bf R}_3$ as well as the points $O$, $A$, $C$, $D$, $M$ and $N$ lie on the plane $x=0$. The point $A$ represents the end point of the body fixed vector ${\bf R}_3$ at this moment. As $\dot{\bf R}_3=[ {\boldsymbol\omega}, {\bf R}_3]$, the point $A$ of the body starts its motion in the direction of the arrow drawn near this point. This determines the directions of motions of other elements as follows. The wheel $D$, with the axis $DO$ fixed at the point $O$, rolls without slipping along the circle drawn on invariable plane. The end of ${\boldsymbol\omega}$ moves on this circle counterclockwise around ${\bf m}$. In the result, all points of the body fixed axis $Dz$ move on the circles of the cone with the axis along the vector ${\bf m}$. In particular, the point $A$ of the body moves on the circle $ACBA$.  

In resume, the axis $Dz$, consisting of the points of the body, generates the surface of the cone, while the remaining points of the body 
instantaneously rotate around this axis. 

The rotation matrix (\ref{s14}) is in correspondence with this picture. Indeed, we observe that it can be decomposed as 
follows: $R(t)=R_{\bf m}(t)\times R_{OZ}(t)$, where $R_{\bf m}(t)$ is the rotation matrix (\ref{s17}), and 
\begin{eqnarray}\label{s19}
R_{OZ}(t)=\left(
\begin{array}{ccc}
\cos \phi t  & -\sin \phi t & 0  \\ 
\sin \phi t &  \cos \phi t & 0 \\
0 & 0 & 1 \\
\end{array}\right), 
\end{eqnarray}
where $\phi=\frac{(I_1-I_3)m_3t}{I_1I_3}$. Then the position ${\bf x}(t)$ of any point of the body at the 
instant $t$ is: ${\bf x}(t)=R_{\bf m}(t)\times R_{OZ}(t){\bf x}(0)$. It is obtained by rotating the initial position vector ${\bf x}(0)$ first around the laboratory axis $OZ$ by the angle $\frac{(I_1-I_3)m_3t}{I_1I_3}$ and then around the ${\bf m}$\,-axis by the angle $\frac{|{\bf m}|t}{I_1}$. 

For completeness, we also present the explicit form for the vector of instantaneous angular velocity of symmetric top
\begin{eqnarray}\label{s20}
{\boldsymbol\omega}(t)=R(t){\boldsymbol\Omega}(t)=(\omega_1, \omega_2, \omega_3)= \qquad \qquad \qquad \qquad \qquad \qquad \qquad \qquad \qquad \qquad  \cr \left( \frac{(I_1-I_3)m_2 m_3}{I_1I_3|{\bf m}|}\sin kt, ~ 
\frac{(I_3m_2^2+I_1m_3^2)m_2}{I_1I_3{\bf m}^2}-\frac{(I_1-I_3)m_2 m_3^2}{I_1I_3{\bf m}^2}\cos kt, ~ 
\frac{(I_3m_2^2+I_1m_3^2)m_3}{I_1I_3{\bf m}^2}+\frac{(I_1-I_3)m_2^2 m_3}{I_1I_3{\bf m}^2}\cos kt \right)
\end{eqnarray}
The magnitude of this vector gives total frequncy of rotation of the body's points at each instant of 
time: ${\boldsymbol\omega}^2=\frac{m_2^2}{I_1^2}+\frac{m_3^2}{I_3^2}$. It is related with two frequences of the rotation matrix as follows: ${\boldsymbol\omega}^2=k^2+\frac{I_1+I_3}{I_1I_3}m_3\phi$.

\begin{acknowledgments}
The work has been supported by the Brazilian foundation CNPq (Conselho Nacional de
Desenvolvimento Cient\'ifico e Tecnol\'ogico - Brasil). 
\end{acknowledgments}

\end{document}